\documentclass{ws-acs}

\begin{document}

\markboth{James Dufty}
{Kinetic Theory and Hydrodynamics for a Low Density Granular Gas}

\catchline

\title{KINETIC THEORY AND HYDRODYNAMICS FOR A LOW DENSITY GRANULAR GAS}

\author{\footnotesize JAMES W. DUFTY}

\address{Department of Physics, University of Florida,\\
Gainesville, FL 32611,USA}

\maketitle

\pub{Received (received date)}{Revised (revised date)}

\begin{abstract}
Many features of real granular fluids under rapid flow are exhibited as well 
by a system of smooth hard spheres with inelastic collisions. 
For such a system, it is tempting to apply standard methods of kinetic theory and 
hydrodynamics to calculate properties of interest. 
The domain of validity for such methods is a priori uncertain due to the inelasticity, but recent systematic studies continue to support the utility of kinetic theory 
and hydrodynamics as both qualitative and quantitative descriptions for 
many physical states. The basis for kinetic theory and hydrodynamic 
descriptions is discussed briefly for the special case of a low density gas.
\end{abstract}

\section{Introduction}	%) A SECTION HEADING

Granular media in rapid, dilute flow exhibit a surprising similarity to
ordinary fluids and the utility of a hydrodynamic description for such
conditions has been recognized for many years \cite{haff83}. This
phenomenology has come under scrutiny in recent years with questions about
the domain of its validity and the associated constitutive equations
appearing in the hydrodynamic equations \cite{kadanoff99,goldhirsh01}.
Answers to such questions can be found in a more fundamental microscopic
description where the tools of nonequilibrium statistical mechanics are
available for critical analysis. An intermediate mesoscopic description
between statistical mechanics and hydrodynamics is that of kinetic theory,
whose applicability to granular matter also poses questions. The objective
in this short communication is to provide an example of a precise context
in which the relevance of kinetic and hydrodynamic descriptions can be
assessed. A more general review of the current status for this problem with
extensive references was prepared for this Workshop \cite{duftytrieste01}
and should be consulted for completeness.

The system considered is a one component gas of $N$ smooth hard spheres at
low density. The inelastic collisions are characterized by a normal
restitution coefficient $\alpha \leq 1$, where $\alpha =1$ is the elastic
limit. The state of the system at each time is specified by a point in the $%
6N$ dimensional phase space and the dynamics is given by uniform motion and
binary inelastic velocity changes at contact. These are the necessary
ingredients to construct a statistical mechanics for this idealized model of
granular media. It is straightforward to write the Liouville equation for
the probability density and from it obtain the BBGKY hierarchy for the
reduced distribution functions \cite%
{brey97statmech,dufty00granada,ernst00rev,vanNoije01sm}. In the next section
the condition of low density is exploited to obtain a complete, closed
kinetic theory description for the gas from the BBGKY hierarchy. The low
density expansion leading to this result is formally the same as that for a
gas with elastic collisions. Consequently, it would appear that the kinetic
description for the granular gas has the same level of validity as that for
normal gases.

\qquad The derivation of hydrodynamics from the kinetic theory is discussed
in the third section. Hydrodynamics is defined generally as the composition
of exact balance equations for the density, energy density (or granular
temperature), and the momentum density (or flow field) plus constitutive
equations for the associated fluxes and cooling rate. Constitutive equations
exist whenever there is a ''normal'' solution to the kinetic theory. Such a
solution can be constructed explicitly for weakly inhomogeneous states,
leading to a Navier-Stokes hydrodynamics for the granular gas \cite%
{brey98boltz,garzo00enskog,garzo01mix}. The need to go beyond the
Navier-Stokes approximation is rare for normal gases, but is not uncommon
for many relevant granular gas states. It remains a challenge for granular
gas theory to understand the form of constitutive equations for these more
general conditions, but this uncertainty should not be interpreted as a
signature of the failure for hydrodynamics to apply.

\section{Kinetic Theory}

Concerns about a kinetic theory description for granular gases are often
based on a too restrictive concept of the prerequisites (e.g., the existence of
an equilibrium state). It is useful therefore to provide a formal
''derivation'' of the kinetic description at low density to demonstate the
close similarity between normal and granular gases, without unwarranted conceptual prejudices.
In this section a small
parameter expansion of the reduced distribution functions is shown to
provide a formal solution to the entire BBGKY \ hierarchy, in parallel to
the corresponding analysis for normal gases \cite{dufty95mirim}%
. The analysis is similar to an expansion proposed by Grad for the hard
sphere gas with elastic collisions \cite{grad58}. In this approach, there is
no reference to concepts such as ''approach to equilibrium'', Maxwellian
distribution, or ''molecular chaos'', and the distinction between inelastic
or elastic collisions plays no explicit role. Thus, superficially at least,
it appears the basis for the low density kinetic theory is the same in both
cases.

The s-particle reduced distribution functions, $f^{(s)}(x_{1},\cdots
,x_{s},t)$, obey the BBGKY hierarchy where $x_{i}=\left( {\bf q}_{i},{\bf v}%
_{i}\right) $. A dimensionless form of this hierarchy is obtained by scaling
the space and time with the mean free path $\ell \equiv 1/(n\sigma ^{2})$
and the mean free time $t_{0}\equiv \ell /v_{0}$. Here, $n$ is the density, $%
\sigma $ is the hard sphere diameter, and $v_{0}$ is some characteristic
velocity. Similarly, the reduced distribution functions are scaled with $%
\left( n/v_{0}^{3}\right) ^{s}$. The resulting dimensionless BBGKY hierarchy
has the form 
\begin{eqnarray}
\lefteqn{\left( \partial _{t}+\sum_{i=1}^{s}v_{i}\cdot \nabla _{i}-\lambda
^{2}\sum_{i<j}^{s}\overline{T}(i,j)\right) f^{(s)}(x_{1},\cdots ,x_{s},t)} 
\nonumber \\
&=&\sum_{i=1}^{s}\int dx_{s+1}\,\overline{T}(i,s+1)f^{(s+1)}(x_{1},\cdots
,x_{s+1},t).  \label{2.1}
\end{eqnarray}%
The binary scattering operator $\overline{T}(i,j)$ for a pair of particles $%
\{i,j\}$ is defined by%
\begin{equation}
\overline{T}(i,j)=\int \ d\hat{\bf \sigma}\ \Theta ({\bf g}_{ij}\cdot 
\hat{\bf \sigma})({\bf g}_{ij}\cdot \hat{\bf \sigma})\left[
\alpha ^{-2}\delta ({\bf q}_{ij}-\lambda \hat{\bf \sigma}%
)b_{ij}^{-1}-\delta ({\bf q}_{ij}+\lambda \hat{\bf \sigma})\right] ,
\label{2.2}
\end{equation}%
where $\hat{{\bf \sigma }}$ is a unit vector along ${\bf q}_{ij}={\bf q}_{i}-%
{\bf q}_{j}$ and $d\hat{{\bf \sigma }}$ denotes a two dimensional solid
angle integration over the sphere for particles at contact. Also, ${\bf g}%
_{ij}\equiv {\bf v}_{i}-{\bf v}_{j}$ is the relative velocity, and $%
b_{ij}^{-1}$ is the scattering operator defined for any function $X({\bf v}%
_{i},{\bf v}_{j})$ by $b_{ij}^{-1}X({\bf v}_{i},{\bf v}_{j})\equiv X({\bf v}%
_{i}^{\prime },{\bf v}_{j}^{\prime }).$ The ''restituting'' velocities are%
\begin{equation}
{\bf v}_{i}^{\prime }\equiv {\bf v}_{i}-\frac{1+\alpha }{2\alpha }({\bf g}%
_{ij}\cdot \hat{{\bf \sigma }})\hat{{\bf \sigma }},\hspace{0.5cm}{\bf v}%
_{j}^{\prime }\equiv {\bf v}_{j}+\frac{1+\alpha }{2\alpha }({\bf g}%
_{ij}\cdot \hat{{\bf \sigma }})\hat{{\bf \sigma }}.  \label{2.4}
\end{equation}%
The $\alpha $ dependence\ of $\overline{T}(i,j)$ contains all aspects of the
inelasticity, and plays no explicit role in the following expansion.

In this dimensionless form the BBGKY \ hierarchy depends on the single
dimensionless parameter $\lambda \equiv \sigma /\ell =n\sigma ^{3}$, the
ratio of the ''force range'' to the mean free path. This parameter is small
at low density, suggesting an expansion for a solution to the entire
hierarchy as a power series in $\lambda $. The dependence on $\lambda $
occurs explicitly as shown on the left side of (\ref{2.1}) and implicitly
through the finite separation of the colliding particles in $\overline{T}%
(i,j)$. The structural features of the expansion in $\lambda $ are simplest
if it is performed at fixed $\overline{T}(i,j)$. The s-particle reduced
distribution functions are taken to have the representation 
\begin{equation}
f^{(s)}(x_{1},\cdots ,x_{s},t)=f_{0}^{(s)}(x_{1},\cdots ,x_{s},t)+\lambda
^{2}f_{1}^{(s)}(x_{1},\cdots ,x_{s},t)+...  \label{2.5}
\end{equation}%
It is then readily shown that the hierarchy is solved exactly to order $%
\lambda ^{2}$ in the form 
\begin{equation}
f_{0}^{(s)}(x_{1},\cdots ,x_{s},t)=\prod_{i=1}^{s}f_{0}^{(1)}(x_{i},t),
\label{2.6}
\end{equation}%
\begin{equation}
f_{1}^{(s)}(x_{1},\cdots ,x_{s},t)=\sum_{j=1}^{s}\prod_{i\neq
j}^{s}f_{0}^{(1)}(x_{i},t)f_{1}^{(1)}(x_{j},t)+\sum_{i<j}^{s}\prod_{k\neq
i,j}^{s}f_{0}^{(1)}(x_{k},t)G(x_{i},x_{j},t),  \label{2.7}
\end{equation}%
where the expression for $f_{1}^{(s)}$ holds for $s\geq 2$. Thus, the
reduced distribution functions for any number of particles are determined as
a sum of products of the single particle functions $f_{0}^{(1)}(x_{1},t)$
and $f_{1}^{(1)}(x_{1},t)$, and pair function $G(x_{1},x_{2},t)$. These
are determined from the set of three fundamental kinetic equations 
\begin{equation}
\left( \frac{\partial }{\partial t}+{\bf v}_{1}\cdot \nabla _{1}\right)
f_{0}^{(1)}(x_{1},t)=J(x_{1},t\mid f_{0}^{(1)}).  \label{2.8}
\end{equation}%
\begin{equation}
\left( \frac{\partial }{\partial t}+{\bf v}_{1}\cdot \nabla _{1}-I_{1}+{\bf v%
}_{2}\cdot \nabla _{2}-I_{2}\right) G(x_{1},x_{2},t)=\overline{T}%
(1,2)f_{0}^{(1)}(x_{1},t)f_{0}^{(1)}(x_{2},t)  \label{2.9}
\end{equation}%
\begin{equation}
\left( \frac{\partial }{\partial t}+{\bf v}_{1}\cdot \nabla
_{1}-I_{1}\right) f_{1}^{(1)}(x_{1},t)=\int dx_{2}\,\overline{T}%
(1,2)G(x_{1},x_{2},t),  \label{2.10}
\end{equation}%
Here $J(x_{1},t\mid f_{0}^{(1)})$ is the Boltzmann-Bogoliubov collision
operator and $I_{1}$, defined over functions of $x_{1}$, is its linearized
form 
\begin{equation}
J(x_{1},t\mid f_{0}^{(1)})=\int dx_{2}\,\overline{T}%
(1,2)f_{0}^{(1)}(x_{1},t)f_{0}^{(1)}(x_{2},t)  \label{2.10a}
\end{equation}%
\begin{equation}
I_{1}h(x_{1})\equiv \int dx_{2}\,\overline{T}(1,2)\left(
f_{0}^{(1)}(x_{1},t)h(x_{2})+h(x_{1})f_{0}^{(1)}(x_{2},t)\right) .
\label{2.11}
\end{equation}

These low density results (\ref{2.8})-(\ref{2.10}) provide the kinetic
theory description for the gas. They are remarkably rich. As expected, the
leading order distribution function $f_{0}^{(1)}$ is the solution to the
Boltzmann equation. The two particle correlations are generated from the
uncorrelated product of Boltzmann solutions through inelastic binary
collisions $\overline{T}(1,2)f_{0}^{(1)}(x_{1},t)f_{0}^{(1)}(x_{2},t)$ on
the right side of (\ref{2.9}). Finally, corrections to the Boltzmann
solution due to correlations are given by a coupling of the distribution
function to the correlations in (\ref{2.10}) (the so-called ''ring''
recollision effects). The solutions and implications of these kinetic
equations can be quite different for elastic and inelastic collisions. But
these differences come from the equations themselves and should not be
interpreted as signatures of their failure to apply. For example, at $\alpha
=1$ a possible solution for an isolated system is $f_{0}^{(1)}(x_{1},t)%
\rightarrow f_{M}(v_{1})$, $G(x_{1},x_{2},t)=0=f_{1}^{(1)}(x_{1},t)$, where $%
f_{M}$ is the Maxwell-Boltzmann distribution. Equation\ (\ref{2.10})
supports $f_{M}$ because energy is conserved, and $G(x_{1},x_{2},t)=0$
because $T(1,2)f_{M}({\bf v}_{1})f_{M}({\bf v}_{2})=0$ for the same reason.
Since energy conservation no longer holds with $\alpha <1$ it is not
surprising that the isolated system does not approach equilibrium, the
Maxwellian is not a stationary solution, and that finite correlations exist.
Indeed, the extent to which such predicted differences agree with
observations from molecular dynamics provide {\em support} for the kinetic
theory, not {\em limitations} on it as is sometimes implied. 

Clearly, the
above derivation has not restricted this kinetic description to isolated
systems or near-equilibrium states. In fact, the most interesting cases of
practical interest are response to boundary conditions and/or external
fields. The similarities between normal and granular fluids is closest for
such ''nonequilibrium'' conditions. Too often, properties of granular gases
are contrasted only to those of the equilibrium state for normal gases.

It is important to note that practical access to the solutions to the above
kinetic equations is possible for a wide range of conditions by direct
simulation Monte Carlo (DSMC) \cite{bird}. Much attention has been given to
the special ''homogeneous cooling state'' (HCS) which is a solution to the
Boltzmann equation with the scaling form%
\begin{equation}
\,f_{0}^{(1)}({\bf v},t)=v_{0}^{-3}(t)n\phi \left( v/v_{0}(t)\right) ,%
\hspace{0.5cm}v_{0}^{2}(t)=2T(t)/m.  \label{2.12}
\end{equation}%
The form of this is HCS distribution is known to good approximation by analytic methods, which
have been confirmed by DSMC \cite{vanNoije01sm}. The correlations for this state also have been
studied in some detail \cite{brey98correl} by both analytic and simulation
methods. Finally, studies of more complex states (e.g., shear flow) also
have been given \cite{brey00model}. In summary, the kinetic theory
description appears to describe well a wealth of new phenomena peculiar to
inelastic collisions.

\section{Hydrodynamics}

Consider now a spatially inhomogeneous state, created either by initial
preparation or by boundary conditions. The local balance equations for the
density $n({\bf r},t)$, granular temperature $T({\bf r},t)$ (or energy
density), and flow velocity ${\bf U}({\bf r},t)$ follow directly by taking
moments of the Boltzmann equation (\ref{2.8}) with respect to $1,$ ${\bf v},$
and $v^{2}$ 
\begin{equation}
D_{t}n+n\nabla \cdot {\bf U}=0,  \label{3.1}
\end{equation}%
\begin{equation}
D_{t}T+\frac{2}{3nk_{B}}\left( P_{ij}\partial _{j}U_{i}+\nabla \cdot {\bf q}%
\right) =-T\zeta ,  \label{3.2}
\end{equation}%
\begin{equation}
D_{t}U_{i}+(mn)^{-1}\partial _{j}P_{ij}=0.  \label{3.3}
\end{equation}%
Here $D_{t}=\partial _{t}+{\bf U}\cdot \nabla $ is the material derivative, $%
P_{ij}({\bf r},t)$ is the pressure tensor and ${\bf q}({\bf r},t)$ is the
heat flux. The form of these balance equations is the same as for fluids
with elastic collisions except for the source term on the on the right side
of (\ref{3.2}) due to the dissipative collisions, where $\zeta \propto
\left( 1-\alpha ^{2}\right) $ is identified as the cooling rate. The fluxes $%
P_{ij}$, ${\bf q}$ \ and the cooling rate $\zeta $ are given as explicit low
degree moments of the distribution function $f^{(1)}(x_{1},t)$ 
\begin{equation}
P_{ij}=\int d{\bf v}\,mV_{i}V_{j}\,f({\bf r},{\bf v},t),\hspace{0.4cm}{\bf q}%
=\int d{\bf v}\,\frac{1}{2}mV^{2}{\bf V}f({\bf r},{\bf v},t),  \label{3.4}
\end{equation}%
\begin{equation}
\zeta =\left( 1-\alpha ^{2}\right) \frac{\beta m\sigma ^{2}}{12nT}\int \,d%
{\bf v}_{1}\,d{\bf v}_{2}\,d\hat{\bf \sigma} \,\Theta (\widehat{\bf
\sigma }\cdot {\bf g})(\widehat{\bf \sigma }\cdot {\bf g})^{3}\ f({\bf r}%
_{1},{\bf v}_{1},t,)f({\bf r}_{2},{\bf v}_{2},t).  \label{3.5}
\end{equation}

The utility of these balance equations is limited without further
specification of $P_{ij}$, ${\bf q}$, and $\zeta $ which, in general, have a
complex space and time\ dependence. However, for a fluid with elastic
collisions this dependence ''simplifies'' on sufficiently large space and
time scales where it is given entirely through a functional dependence on
the fields $n$, $T$, and ${\bf U}$. The resulting functional dependencies of 
$P_{ij}$ and ${\bf q}$ on these fields are called constitutive equations and
their discovery can be a difficult many-body problem. The above balance
equations, together with the constitutive equations, become \ a closed set
of equations for $n$, $T$, and ${\bf U}$ called {\em hydrodynamic equations}%
. This is the most general and abstract notion of hydrodynamics, which
encompasses both the Navier-Stokes form for small spatial gradients and more
general forms for nonlinear rheological transport. The primary feature of a
hydrodynamic description is the reduction of the description from many
microscopic degrees of freedom to a set of equations for only five local
fields.

The critical problem for a hydrodynamic description is therefore to
determine the existence and form of the constitutive equations. It is clear from (\ref{3.3})
and (\ref{3.4}) that they can be obtained if the Boltzmann equation admits
a ''normal'' solution, whose space and time dependence occurs entirely
through its functional dependence on the fields 
\begin{equation}
f({\bf r},{\bf v},t)=F\left( {\bf v\mid }n,T,{\bf U}\right).  \label{3.6}
\end{equation}%
The fluxes and cooling rate then inherit this space and time dependence and
become constitutive equations. The space and time dependence of the fields
follows from solution to the resulting hydrodynamic equations to complete
the self-consistent description of $F$. An example is given by the HCS
distribution in (\ref{2.12}) where there is no space dependence and all of
the time dependence occurs through $T(t)$. The latter is determined from the
hydrodynamic equation (\ref{3.2}), which reduces to $\partial _{t}T=$ $%
-T\zeta $. Use of (\ref{2.12}) in (\ref{3.5}) gives the constitutive
equation $\zeta =\zeta _{0}T^{1/2}$ where $\zeta _{0}$ is a constant.

For gases with elastic collisions the prototypical hydrodynamics is that of
the Navier-Stokes equations. The corresponding constitutive equations,
Newton's viscosity law and Fourier's heat law, follow from a normal solution
to the Boltzmann equation obtained from an expansion in the spatial
gradients. The reference
state is the {\em local} Maxwellian whose parameters are the exact density,
temperature, and flow velocity for the nonequilibrium state being described.
Deviations from this reference state are proportional to the spatial
gradients of the temperature and flow field. The systematic expansion for
the normal solution to the Boltzmann equation is generated by the
Chapman-Enskog method. There have been three main objections/reservations/concerns
regarding application of this method for granular gases: 1) the
absence of an equilibrium state as the basis for the local Maxwellian
reference state, 2) the inherent time dependence of any reference
state due to collisional cooling, and 3) the inclusion of the
energy (temperature) as a hydrodynamic field when it is not associated with
a conserved density and does not have a time scale solely characterized by
the degree of spatial inhomogeneity.

The first two concerns are primarily technical rather than conceptual issues that can
be answered by direct application of the Chapman-Enskog method to see if it
indeed generates a normal solution to the granular Boltzmann equation. 
Consider a state for which the spatial variations of $n$, $T$, and ${\bf U}$
are small on the scale of the mean free path, $\ell \nabla \ln n<<1$, where $%
\ell =1/n\sigma ^{2}$ is the mean free path. Then it is expected that the
functional dependence of the normal solution on the hydrodynamic fields can
be made local in space\ through a Taylor series expansion about a point $%
{\bf r}$ for which the distribution function is being
evaluated. Let 
$\epsilon $ denote a formal small ''uniformity'' parameter measuring the
small spatial gradients in the fields. It is worth emphasizing that there is
no restriction on the cooling rate $\zeta $; only the spatial gradients are
being ordered by the uniformity parameter. The distribution function,
collision operator, and time derivatives are given by the representations 
\begin{equation}
F=F^{(0)}+\epsilon F^{(1)}+\cdots ,\hspace{0.2in}J_{E}=J^{(0)}+\epsilon
J^{(1)}+\cdots ,\hspace{0.2in}\partial _{t}=\partial _{t}^{(0)}+\epsilon
\partial _{t}^{(1)}+...  \label{3.7}
\end{equation}%
The coefficients in the time derivative expansion are identified from the
balance equations with a similar expansion for $P_{ij}$, ${\bf q}$, and $%
\zeta $ generated through their definitions (\ref{3.4}) and (\ref{3.5}) as
functionals of $F$. As for elastic collisions the leading term $F^{(0)}$ is
further constrained to have the same moments with respect to $1,v^{2},$ and $%
{\bf v}$ as the full distribution $F$. This assures that the hydrodynamic
fields occuring in $F^{(0)}$ are exactly those for the nonequilibrium
state, and hence $F^{(0)}$ is normal. To zeroth order in $\epsilon $ the Boltzmann kinetic equation becomes
\begin{equation}
\frac{1}{2}\zeta ^{(0)}{\bf \nabla}_{V}{\bf \cdot }\left( {\bf V}%
F^{(0)}\right) =J^{(0)}[F^{(0)},F^{(0)}].  \label{3.8}
\end{equation}%
where ${\bf V}={\bf v-U}$. 

The first concern above regarding
the reference distribution now can be addressed. The distribution $F^{(0)}$ 
is not free to be chosen, but rather is determined
by the kinetic equation itself. For the granular gas it is {\em not}
the local Maxwellian, but rather the local (normal) HCS solution (i.e., (\ref{2.12})
with the density, temperature, and flow field replaced by their
nonequlibrium values). There is no {\it a priori} requirement of an
equilibrium state for the Chapman-Enskog method to apply, and in fact early
applications based on a local equilibrium state are inconsistent and lead to incorrect transport
coefficients even for weak dissipation. The concept of "approach to equilibrium" is 
no longer relevant for granular gases.

Mathematically, the changes in this method for granular gases arise from the
fact that the time derivative of the temperature does not vanish to lowest
order in $\epsilon $, as it does for a gas with elastic collisions. The 
reference state $F^{(0)}$ incorporates this zeroth order time dependence of the
temperature even for strong dissipation. This is the origin of the second concern
noted above. However, since $F^{(0)}$ is
normal, it necessarily has the exact time dependence of all hydrodynamic fields. 
The only difference for granular
gases is the introduction of a new time scale $1/\zeta ^{(0)}$ in the
reference state. There is nothing {\em a priori} inconsistent with
a description of slow spatial decay towards a time dependent reference
state. In fact, this is an important feature of the Chapman-Enskog scheme for both granular and
normal fluids alike - use of a time independent reference state would limit
the derivation to only {\em linear} hydrodynamics.

Implementation of the Chapman-Enskog method to the first order in $\epsilon $
is now straightforward and has been carried out in detail and without
approximation recently for the Boltzmann equation \cite{brey98boltz} (and for
its dense fluid generalization, the Enskog equation \cite{garzo00enskog}); the case of a two component
mixture is considered in \cite{garzo01mix}. The constitutive equations for
the one component fluid found to this order are%
\begin{equation}
P_{ij}\rightarrow p\delta _{ij}-\eta \left( \partial _{j}U_{i}+\partial
_{i}U_{j}-\frac{2}{3}\delta _{ij}\nabla \cdot {\bf U}\right) -\gamma \delta
_{ij}\nabla \cdot {\bf U},\hspace{0.3in}  \label{3.9}
\end{equation}%
\begin{equation}
{\bf q}\rightarrow -\kappa \nabla T-\mu \nabla n,  \label{3.10}
\end{equation}%
The form of the pressure tensor is the same as that for fluids with elastic
collisions, where $\eta \left(\alpha \right) $ is the shear viscosity depending on
the restitution coefficient. 
The heat flux is similar to Fourier's law, where $\kappa \left(\alpha
\right) $ is the thermal conductivity. There is a new transport
coefficient $\mu \left(\alpha \right) $ coupling the heat flux to a
density gradient, which vanishes at $\alpha =1$. \ The transport coefficients
are given in terms of solutions to inhomogeneous integral equations
involving the linearized Boltzmann collision operator. Solubility conditions for the
existence of solutions have been proven and approximate solutions in terms
of polynomial expansions have been obtained for practical purposes, just as
for the case of elastic collisions \cite{garzo00enskog}. Consequently, the
transport coefficients are known as explicit functions of $\alpha$. 
The Chapman-Enskog method places
no explicit restriction on $\alpha $ so the results are not limited
to weak dissipation.
	
It remains to discuss the third concern, use of the temperature as a hydrodynamic field which is
no longer associated with a conserved density. 
Analysis of the Navier-Stokes approximation described above shows there are two classes of 
inherent time scales
to this hydrodynammic description. The first scales with the wavelength of the phenomena considered
and can be made large (long time scales) by considering 
sufficiently smooth disturbances. This is the case for gases with elastic collisions. 
For granular gases there is
a second time scale in the temperature equation, the inverse cooling rate, which is bounded for a given value
of the restitution coefficient. The concern is that this new time scale may be shorter than that required for
validity of the normal solution to the kinetic equation. 

Qualitatively,
a wide class of solutions for spatially inhomogeneous states evolves
in two stages. During a short transient period of the order of the mean free
time (the kinetic stage), the velocity distribution approaches closely the
local distribution $F^{(0)}$ and becomes normal as in (\ref{3.6}). Subsequently, on a
longer time scale the space and time dependence of the distribution occurs
only through the fields that are governed by hydrodynamic equations. This notion of 
a two stage relaxation is similar to that for gases with elastic collisions 
and is confirmed for granular gases by DSMC. The question, therefore, is whether the inverse cooling rate
is smaller than the mean free time. Certainly, this is the case for weak dissipation
since $\zeta ^{(0)}$ is proportional to $\left( 1-\alpha
^{2}\right) $. For strong dissipation it is necessary to study in detail the spectrum
of non hydrodynamic (kinetic) modes to find the slowest mode and compare
it with the inverse cooling rate. This has been done for the diverse examples of 
Brownian motion, kinetic models, and uniform shear flow (see \cite{duftytrieste01}). 
In all cases it is found that there is a 
separation of time scales between the slowest kinetic mode and the inverse cooling rate,
even at strong dissipation. Based on these examples, it appears that inclusion of the temperature
in the hydrodynamic description is justified for a wide range of degrees of dissipation.  

The validity of Navier-Stokes hydrodynamics and the dependence of the
transport coefficients on the restitution coefficient has been verified in a
number of simulations, both DSMC and MD, giving excellent agreement with the
predictions from the Chapman-Enskog method. The tests at low density based
on the Boltzmann equation have been reviewed recently by Brey and Cubero \cite%
{brey01rev}. 

The Navier-Stokes equations equations associated with (\ref{3.9}) and (\ref{3.10}) are
also known as Newtonian hydrodynamics.  Although the
Chapman-Enskog method can be carried out to second order in $\epsilon $
(Burnett order), it is likely that failure of the Navier-Stokes
approximation signals more complex non Newtonian behavior for which 
other methods to construct the normal
solution are required that are not based on a small gradient expansion. 

\section{Discussion}

Two questions have been addressed here: 1) can kinetic theory provide a
valid mesoscopic description of rapid flow (fluidized) granular media?, and
2) can a hydrodynamic description be formulated and justified for a
macroscopic description? The origin of the kinetic description for a low density
gas has been shown to follow from an expansion that makes no reference as to 
whether the collisions are elastic or inelastic. The latter affects only the 
solution to these equations. It appears
that many of the questions, concerns, and objections raised regarding the validity
of a kinetic theory description are removed in this way. Furthermore, kinetic theory
provides a powerful tool for analysis and predictions of
rapid flow gas dynamics when combined with the method of DSMC for numerical solution. 
Its full potential for both conceptual and practical questions has not yet been exploited.

An important use of the kinetic theory is to determine if a 
hydrodynamic description applies and to define its domain of validity. Equations (\ref{2.7}) - (\ref{2.9}) leave
no room for speculation. For given initial and boundary conditions the solution either approaches a
normal form on some space and time scale or it does not. The issue of a hydrodynamic
description is now a precise mathematical question. This is particularly important
if the hydrodynamic description is more complex than that of Navier-Stokes.  
Non Newtonian
hydrodynamics is rare or unphysical for simple atomic systems with elastic collisions. In
contrast, they it is more common for granular gases in steady states where
the gradients are strongly correlated to the coefficient of restitution. Kinetic theory
is perhaps the only systematic means to determine the constitutive equations in these cases.

It is clear from recent studies that granular media
exhibit a wide range of interesting phenomena for which a Navier-Stokes
hydrodynamics is an accurate and practical tool (see \cite{duftytrieste01} for specific
examples and references). Although the context here has been limited to low density
gases a similar kinetic theory basis has been developed for Navier-Stokes hydrodynamics in moderately dense 
fluids  \cite%
{garzo00enskog}. More generally, non Newtonian behavior puts granular fluids in a class of
complex materials with mysterious, as yet unexplained, properties \cite{kadanoff99,goldhirsh01}.
Kinetic theory and hydrodynamics (in the broader sense) can be
expected to provide much of this explanation.

\section*{Acknowledgements}

This research was supported in part by National Science Foundation grant PHY 9722133.

\end{document}